\documentclass[fleqn,a4]{article}

\usepackage{sandro}
\usepackage{amssymb}

\newcommand{\startclauses}{ \vspace{3mm} \\ \begin{math}
\begin{array}{llllll}}
\newcommand{\stopclauses}{\end{array} \end{math}
\vspace{3mm} \\}

\newcommand{\TEXC}[1]{\mbox{$ \exists X C $}}

\newcommand{\TEYD}[1]{\mbox{$ \exists Y D $}}

\newcommand{\ES}{\mbox{$\emptyset$}}    

\newcommand{\la}{\ensuremath{\:\leftarrow\:}}
\newcommand{\ra}{\ensuremath{\:\rightarrow\:}}

\newcommand{\A}{\ensuremath{\ \wedge\ }}

\newcommand{\LL}{\ensuremath{\ldots}}

\newcommand{\NI}{\noindent}
\newcommand{\HB}{\hfill{$\square$}}
  
\newcommand{\II}{\vspace{2 mm}}

\newcommand{\ol}[1]{\ensuremath{\tilde{#1}}}

\newcommand{\var}{{\it Var}}

\def\smallromani{\renewcommand{\theenumi}{\roman{enumi}}
\renewcommand{\labelenumi}{(\theenumi)}}

\newenvironment{program}{\tt \begin{tabbing}pro\= {\tt pro}\= clause \kill}{\end{tabbing}}

\newcounter{symbol}
\newcommand{\indexsyma}[1]%
{\stepcounter{symbol}\index{zzz1 \thesymbol @\protect#1}}
\newcommand{\indexsymb}[1]%
{\stepcounter{symbol}\index{zzz2 \thesymbol @\protect#1}}
\newcommand{\indexsymc}[1]%
{\stepcounter{symbol}\index{zzz3 \thesymbol @\protect#1}}
\newcommand{\indexsymd}[1]%
{\stepcounter{symbol}\index{zzz4 \thesymbol @\protect#1}}
\newcommand{\indexsyme}[1]%
{\stepcounter{symbol}\index{zzz5 \thesymbol @\protect#1}}

\usepackage{alltt}
\newcounter{jon}

\newcommand{\todoloo}{\ifthenelse{\value{jon}=0}{}{
{ {\huge \textsf{There are still }\arabic{jon}%
\textsf{comments by Jon}}} }} 

\newenvironment{mirprogram}{\sf \begin{tabbing}{\tt pro}\= {\tt pro}\=
    {\tt prolungamentissimoo }\=
    \hspace{5cm} \=
    clause \kill}{\end{tabbing}}

\newenvironment{haskell}
{\sf \begin{tabbing}{\tt pro}\= 
    {\tt proprolungamentissimoo }\= {\tt pro}\= {\tt pro}\=
    clause \kill}{\end{tabbing}}

\newcommand{\PP}{{\tt P}}

\renewcommand{\ol}[1]{\ensuremath{\overline{#1}}}
\newcommand{\ols}[1]{\ensuremath{\overline{\sf #1}}}
\newcommand{\olt}[1]{\ensuremath{\overline{\tt #1}}}

\title{
The (Lazy) Functional Side of Logic Programming}
\author{Sandro Etalle\\ 
\small University of Maastricht\\
\small P.O.\ Box 616, 6200 MD Maastricht\\
\small The Netherlands\\
\emph{\small etalle@cs.unimaas.nl} 
\and Jon Mountjoy\\
\small University of  Amsterdam\\
\small Kruislaan 403, 1098 SJ Amsterdam\\
\small The Netherlands\\
\emph{\small jon@wins.uva.nl}}
\date{}

\begin{document}
\pagestyle{plain}
\newtheorem{sander}{\emph{Sandro's stuff}}
\newcommand{\san}[1]{\begin{sander}\em #1 \end{sander}}

\maketitle
\begin{abstract}
  The possibility of translating logic programs into functional ones
  has long been a subject of investigation.  Common to the many
  approaches is that the original logic program, in order to be
  translated, needs to be \emph{well-moded} and this has led to the
  common understanding that these programs can be considered to be the
  ``functional part'' of logic programs.  As a consequence of this it
  has become widely accepted that ``complex'' logical variables, the
  possibility of a dynamic selection rule, and general properties of
  non-well-moded programs are exclusive features of logic programs.

This is not quite true, as some of these features are naturally found
in \emph{lazy} functional languages. 

We readdress the old question of what features are exclusive to the
logic programming paradigm by defining a simple translation applicable
to a wider range of logic programs, and demonstrate that the current
circumscription is unreasonably restrictive.
\II


\noindent
\emph{ACM Computing Classification System:} D.1.1 (applicative --
functional -- programming); D.1.6 (logic programming); D.3.2 (language
classifications) F.3.3 (studies of program constructs) programs)

\end{abstract}
\thispagestyle{empty}

\section{Introduction}
The possibility of translating logic programs into functional ones has
long been a subject of investigation. Among the different proposals
\cite{Mar94-trs,Mar95,GW92,kRKS98,Red84,vRaa97}.  Such systems are
usually devised for one of the following purposes: for proving program
properties, for providing better insight on the relation between
functional and logic languages, or -- to a minor extent -- for
improving program performance.

Common to all the approaches mentioned is that the original logic
program, in order to be translated, needs to be \emph{well-moded} and
this has led to the common understanding that these programs can be
considered to be the ``functional part'' of logic programs. This is
confirmed by the following statement in \cite{Mar95}: ``$\ldots$ the
class of functionally moded (well-moded and simply moded) programs can
be rightly considered \emph{the} functional core of logic programs''.

Well-moded programs have, among other features, a straightforward
left-to-right dataflow model (see \cite{AE93,AM94}) and prohibit the
use of \emph{logical variables} to their full potential such as in
complex logical data structures like difference-lists. As a
consequence of this it is now widely accepted that ``complex'' logical
variables, the possibility of a dynamic selection rule, and general
properties of non-well-moded programs are exclusive features of logic
programs.

This is not quite right. At least, not to the extent that one is
brought to think.

In this paper we show, among other things, that logical structures
such as difference lists have a natural counterpart in \emph{lazy}
functional programs; i.e. that most programs using difference-lists
are functional in nature. This shows immediately that many common
non-well-moded programs are functional in nature and that
well-modedness is thus not a necessary attribute of those logic
programs behaving functionally. We do this by employing a
straightforward -- literal -- translation of moded logic programs into
Haskell, a lazy functional language.

Furthermore, we use the same translation system to show that some
programs requiring a dynamic scheduling mechanism are also
intrinsically functional.

Summarizing, in this paper we readdress the old question of what
features are exclusive to the logic programming paradigm and
demonstrate that the current circumscription is unreasonably restrictive.


\section{Preliminaries}

Due to space constraints we omit preliminaries and assume that the
reader is acquainted with the terminology and the main results of
logic programming theory (see \cite{Apt90,Llo87}). In this paper we
use over-lined characters to indicate (a possibly empty) sequence of
objects,
so $\ol t$ can denote a sequence $t_1,\ldots, t_n$ of terms, $\ol x$ a
sequence of variables and $\olt A$ a sequence of atoms (i.e.\ a
query).  To avoid confusion with built-in symbols, we use $\equiv$ to
indicate syntactic equivalence.  

In what follows we study logic programs executed by means of the {\em
  LD-resolution\/}, which consists of the SLD-resolution combined with
the leftmost selection rule. An SLD-derivation in which the leftmost
selection rule is used is called an {\em LD-derivation}.

\subsection{Modes for Logic Programs}

This section is partially borrowed from \cite{AE93}, we refer to the
appendix and to \cite{AM94} for further information over well-moded
logic programs.

\begin{definition}
\label{def:mode}
  Consider an $n$-ary relation symbol $\tt p$. By a \emph{mode} for
  $\tt p$ we mean a function $\tt m_p$ from $\{{1, \LL, n}\}$ to the
  set $\tt \{{\tt In, Out}\}$. If $\tt m_p(i) = \texttt{In}$, we call $i$
  an \emph{input position} of $p$ and if $\tt m_p(i) = \texttt{Out}$,
  we call $i$ an \emph{output position} of $\tt p$ (both w.r.t.
  $\tt m_p$).  \HB
\end{definition}

\NI An $n$-ary relation ${\tt p}$ with a mode $\tt m_p$ will be
denoted by $ \tt p(m_p(1), \LL , m_p(n)).  $ For example, for programs
\texttt{member} and \texttt{append} we typically have the following
listings and  modes:
\begin{program}
\> mode member(In, In).\\[2mm]
\> member(El, [El|\_]).\\
\> member(El, [\_|Rest]) \la member(El, Rest).\\[2mm]
\> mode append(In, In, Out).\\[2mm]
\> append([ ], \_, [ ]).\\
\> append([H|Tail], List, [H|Tail']) \la append(Tail, List, Tail').
\end{program}
Modes indicate how the arguments of a relation should be used.  We
assume that to each relation symbol is associated a unique mode.
Multiple modes can be obtained by simply renaming the relations.

In presence of modes, we require the programs and the queries to be
somehow naturally \emph{consistent} wrt them. Before we introduce the
notion of consistency we have to provide some further notation.  When
writing an atom as $\tt p(\olt{u},\ \olt{v})$ we now assume that
$\olt{u}$ is a sequence of terms filling in the input positions of
$\tt p$ and $\olt{v}$ is a sequence of terms filling in the output
positions. Thus, for notational simplicity, we assume that the input
positions come first.

Let us call \emph{producing} the input position of the head and the
output positions of the body atoms, and \emph{consuming} the other
positions of a clause, we have the following definition

\begin{definition}[Consistent]
  A clause (query) is \emph{consistent} iff every variable occurs in
  at least one producing position. \HB
\end{definition}

The last LP notion we need is the one of \emph{plain} program. Here
and in the sequel, a set of terms is called \emph{linear} if every
variable occurs at most once in it. In other words, a sat of terms is
linear iff no variable has two distinct occurrences in any of the
terms and no two terms have a variable in common.

\begin{definition}[Plain] \label{def:plain} 
  A clause $\tt p_0(\olt s_0,\;\olt t_{n+1}) \la
  p_1(\olt s_1,\;\olt t_1), \LL, p_n(\olt s_n,\;\olt t_n)$ is
  called \emph{plain} if 
\begin{enumerate}\smallromani
\item $\olt t_1, \dots , \olt t_n$ is a linear family of variables;
\item $\olt s_0$ is linear.
\end{enumerate}
\NI A query \olt{Q} is called \emph{plain} iff the clause
$\tt q \la \olt Q$ is, where \texttt{q} is any (dummy) atom of zero
arity.  \II

\NI
A program is called \emph{plain} if every clause of it
  is.  \HB
\end{definition}

Thus a plain program is a program in which producing positions are
filled in by variables and in which a variable occurs in at most one
producing position. 

Condition (i) is similar to, though less restrictive, than the one of
simply moded programs as defined in~\cite{AE93} (as we do not impose
an ordering constraint).

Our translation requires programs to be consistent and plain.  This is
far less restrictive than well-modedness plus simple-modedness, and as
we shall see allows us to capture a broader segment of functional
behaviour found in logic programs.  Indeed, it is now perhaps
\emph{too} lenient, but suffices for the goals of this paper to
broaden the characterization of logic programs. Regarding the (non)
restrictiveness of the concepts of plain and consistent programs, we
have the following:

\begin{remark}
\label{rem:sm}
It is important to realize that most programs are plain\footnote{this
  assertion is substantiated by the fact that most programs are simply
  moded, as shown by ``mini-survey'' at the end of \cite{AE93}.}, and
that non plain programs can naturally be transformed into equivalent
plain ones, virtually all consistent programs are either plain or
safely translatable into a plain form.  This is also practically
demonstrated by the fact that the language Mercury employs a
pre-processing phase in which all programs are translated into a
\emph{superheterogeneous} form (which is very similar to the form of
plain programs).  Concerning (ii), we can always transform a
consistent program $\tt P$ into an equivalent consistent program $\tt
P'$ which satisfies it.  For instance, for the {\tt member} program
defined above, we can transform its first clause into {\tt member(El,
  [Head|Rest]) \la El == Head}.  It is also worth noticing that {\tt
  append} is already input-linear.  \HB
\end{remark}

\subsection{Haskell Programs}

Our translation system maps logic programs into lazy functional
programs, which are written in (a subset of) Haskell~\cite{hudak5}.
The subset we use includes the proposed extension of \emph{pattern
  guards}~\cite{pattern}, which we describe below.

The programs we are going to generate are built as sets of
\emph{equations}, each of the following form:
\begin{mirprogram}
\> f s$_1$ $\ldots$ s$_j$ \> \> $|$ guard$_{1,1}$, $\ldots$,
                                    guard$_{1,j}$ \> = result$_1$\\
\> \> \> $\vdots$\\
\> \> \> $|$ guard$_{n,1}$, $\ldots$, guard$_{n,m}$ \> = result$_{n,1}$ \\
\> \> \> $|$ otherwise \> = result$_{n+1}$
\end{mirprogram}

where \textsf{f} is a function symbol, \textsf{s$_1$ $\ldots$ s$_j$}
are \emph{parameters} and $\sf guard_{x,y},\ otherwise$ are \emph{guard
  qualifiers}. The `$|$' introduces a guard, and the `$,$' acts as a
logical conjunctive. Pattern matching may take place on the
parameters.  Without pattern guards, the guard qualifiers would have
to be boolean expressions; that is, we would only return $\sf
result_1$ if the associated guards $\sf guard_{1,1} \ldots
guard_{1,j}$ all evaluate to true. 
The semantics of Haskell dictates that definitions and guards are
tried in sequential order.

The situation with pattern guards is somewhat different.
In fact, \emph{patterns guards} can also contain
\emph{let-expressions} (which are defined as usual) and
\emph{pattern-matching expressions} which are expressions of the form
\textsf{pattern \la term}, and whose semantics is the following: if
\textsf{term} matches with \textsf{pattern} then the variables in
\textsf{pattern} are appropriately instantiated, and the
pattern-matching guard returns true, otherwise it return false.
Consider:
\begin{mirprogram}
\> f z  \> $|$ let x = g z \\
\>      \> $,$ [y] \la x \\
\>      \> $,$ y $\geq$ 10 \> = (True,1) \\
\>      \> $|$ otherwise \> = (False,0)
\end{mirprogram}
Here, the tuple \textsf{(True,1)} will only be returned if the two
qualifiers in the first guard succeed.  That is, 1) the argument
\textsf{x} can be pattern matched to a list of one element (denoted by
\textsf{[y]}, which also binds \textsf{y} to this element), and 2) the
boolean condition \textsf{y $\geq$ 10} is true. In all this, the value
of $\sf x$ is determined by $\sf g\ z$.  If any of these fail, then
the second guard is tried.  In this case, the special guard
\textsf{otherwise} will be tried, which always succeeds.  A
\textsf{let} qualifier can also be introduce recursive bindings; this
will become clear in the sequel. A more detailed example
is presented in Appendix \ref{app:more}.

As explained below, we need to capture the fact that a predicate might
succeed (possibly returning a computed answer substitution), or fail.
To do so, we introduce a new datatype $\sf Result$ in our Haskell
programs by:

\begin{mirprogram}
\> data Result $\alpha$ = Suc $\alpha$ $|$ Fail \> 
\end{mirprogram}

That is, the datatype $\sf Result$ has two constructors, $\sf Fail$
and $\sf Suc$, the latter of which can be applied to some term.  

Note that the Haskell programs which we generate are not the obvious
programs that a functional programmer would write -- this is not the
intention.  They do, however, do what the logic programmer intended.
All of the programs given in this paper can be compiled by any Haskell
compiler supporting pattern guards. 

We also want to mention that despite the fact that pattern guards are
 an important feature of our translation, we could do without them --
 at the price of less elegant translation.  A Haskell compiler will
 usually regard these as syntactic sugar anyway, and compile them into
 more basic primitives already found in Haskell.  This implies that
 all the statements we are going to give in the sequel are true
 regardless of the availability of a pattern guard construct in the
 target language.

\section{A Translation System}

In logic programming, queries can succeed, loop or fail. This third
possibility is of crucial importance, since it is often used as a
control mechanism. As an example, one can consider the following
programming scheme:
\begin{program}
\>  p(X) \la generate(X), test(X).
\end{program}
Where \texttt{test} verifies that the value produced by \texttt{generate}
is appropriate, and failure and backtracking take care of the
ill-formed terms.  Another common scheme is the following one:
\begin{program}
\> p(X) \la test\_a, X = 1.\\
\> p(X) \la test\_b, X = 2.
\end{program}
Where $\texttt{test\_a}$ and $\texttt{test\_b}$ model a typical
\texttt{case} statement, and the selection of the right branch is done
via the failure and backtracking mechanism.  \II

Nevertheless, relations which are ``not supposed to fail'' are quite
common in logic programming. We say that a relation is ``not supposed
to fail'' if -- when called in a ``correct'' way -- produces at least
one answer. Examples of such relations are \texttt{sort},
\texttt{flatten} and \texttt{append}, (this latter, for calls of the
form $\mathtt{append(l_1,\ l_2,\ X)}$, where $\tt l_1$ and $\tt l_2$
are lists, and $\tt X$ is a new variable will always produce one
answer).

The ubiquity of predicates which are not supposed to fail is confirmed
by the fact that Mercury requires the programmer to specify for each
relation symbol, whether it might fail or not. This information is
then used to generate  optimized code.

We do the same thing for our translation, and from now on we assume
that the set of predicate symbols is partitioned into two disjoint
sets, namely
\begin{description}
\item[non-test] predicates, which, when called, are expected to
  produce at least one answer (such as {\tt append}), and
\item[test] predicates, which when called are allowed to report no
  answer, i.e., to fail immediately or succeed (such as {\tt member}
  and $\tt <$).
\end{description}

\newcommand{\QQ}{\ensuremath{{\cal Q}}}

Thus, we have the following definition.
\begin{definition}
\label{def:partitioning} A \emph{partitioning} is a map from 
the set of predicate symbols into the set $\{ \emph{test},
\emph{non-test} \}$. 
\II

Let $\tt P$ be a program and \QQ\ be a set of queries, we say that
$\tt P$ is \emph{correct} wrt. \QQ\ iff for every $\olt A \in \QQ$
every, time that a non-test atom $\tt B$ is selected in a
LD-derivation of $\olt A$ in $\tt P$ then $\tt B$ has at least one
successful 
LD-derivation.  \HB
\end{definition}

Thus every program is correct wrt. the trivial partitioning in which
all predicates are \emph{test}. Checking correctness is orthogonal to
the purposes of this paper, but we should mention that it can be done
either using abstract interpretation \cite{DGH97} or on modes and
types \cite{PR97}; also Mercury employs a system based on modes and
types in order to check that the programs are consistent (modulo
non-termination) wrt. the partitioning provided by the programmer.

The partition into test and non-test predicates exposes the implicit
failure mechanism present in logic programs.  Our translation will
transform non-test predicates as ordinary functions, but transform
test predicates by returning something of the type \textsf{Result$\ 
  \alpha$}, allowing us to indicate failure.  Essentially if a
function fails in the logic programming sense, then a value of $\sf
Fail$ will be returned.  Each ``value'' returned from a test predicate
is only every used in a function if it is successfully matched against
$\sf Succ\ a$ indicating that no failure occurred(it was not $\sf
Fail$).  The combination of plain programs with a partition makes it
easy to identify the logic programs that can be mapped to functions,
and which functions to enhance by mimicking the implicit failure
mechanism.  Now, let $\tt p$ be a predicate symbol with mode
\[\tt p(\ \overbrace{\tt In,\ldots,In}^{\it i},\  
\overbrace{\tt Out,\ldots, Out}^{\it j}\  )\] 
Then $\tt p$ can naturally be translated into a function of type
\[
\begin{array}{ll}
\sf p: T_1 \times \cdots \times T_i \ \ra \ (S_1 \times \cdots
  \times S_j)& \hbox{if $\tt p$ is a \emph{non-test} predicate}\\
\sf p: T_1 \times \cdots \times T_i \ \ra \ Result ( S_1 \times \cdots
\times S_j)  & \hbox{if $\tt p$ is a \emph{test} predicate}
\end{array}
\]
Where $\sf T_i$ and $\sf S_i$ are appropriate Haskell types. Here we
will not bother further with the type that the translated predicate
has: the Haskell compiler will be able to infer it autonomously; what
it is important to see is that the Haskell counterpart of \texttt{p}
is a function which maps a tuple with $i$ elements into a tuple
containing $j$ elements, possibly embedded in the $\sf Return$
datatype depending on whether \texttt{p} is a non-test predicate or
not.  We shall employ the value $\sf Fail$ to denote the functional
counterpart of failure\footnote{Confusingly, if a test predicate
  returns no values then we will return $\sf Suc\ ()$, where $()$
  looks like an empty tuple but which is in actual fact the only
  element in the unit type.}.


\subsection{The Translation}
\label{sec:plain}

Our translation method requires the program to be translated to be
\emph{consistent} and \emph{plain}. Of these conditions,
\emph{consistency} is the only crucial one, in fact as stated in
Remark \ref{rem:sm} it is (virtually) always possible to transform a
consistent program into an equivalent program which is plain;
moreover, most programs are plain already.

Now, we can transform the logic program into a Haskell one via a
simple syntactic transformation. First, we have to translate
variables, terms and predicate symbols; this is done in a
straightforward way: one just has to respect the syntactic conventions of
the two languages (uppercase and lowercases, and built-in predicates).
Of course, predicate symbols are transformed into non-constructor
function symbols.  In the sequel we use sans-serif characters for
Haskell constructs and typewriter font for logic programming ones, for
instance, $\sf t, s$ denote the Haskell counterpart of the LP terms
$\tt t, s$.

\begin{definition}[Translation]\label{def:translation}
  Let $\PP$ be a logic program, and \\[2mm] 
$
\begin{array}{lccccccc}
\tt p(\olt t_1, \olt s_1) &\la& 
\tt p_{1,1}(\olt i_{1,1},\olt o_{1,1}), \ldots,
\tt p_{1,k_1}(\olt i_{1,k_1}, \olt o_{1,k_1}),\ 
\tt q_{1,1}(\olt u_{1,1}, \olt v_{1,1}), \ldots,
\tt q_{1,l_1}(\olt u_{1,l_1}, \olt v_{1,l_1})
.\\
& \vdots\\  
\tt p(\olt t_n, \olt s_n) &\la& 
\tt p_{n,1}(\olt i_{n,1}, \olt o_{n,1}), \ldots,
\tt p_{n,k_n}(\olt i_{n,k_n}, \olt o_{n,k_n}),\ 
\tt q_{n,1}(\olt u_{n,1}, \olt v_{n,1}), \ldots,
\tt q_{n,l_n}(\olt u_{n,l_n}, \olt v_{n,l_n}).
\end{array}
$\\[2mm] 
be the set of clauses of $\PP$ defining predicate \texttt{p},
where the predicates $\tt p_{i,j}$ are test predicates and the
predicates $\tt q_{i,j}$ are the non-test ones.  Here we assume that
the clauses had been renamed apart, i.e., that they share no
variables.
\begin{itemize}
\item If $\tt p$ is a \emph{test} predicate, then the
  \emph{translation} of the above section into Haskell is the
  following script (for the moment the underlined parts have to be
  treated as if the underline wasn't there): 

{\sf
\begin{tabbing}
{\tt pro1}\= {\tt pro1}\= {\tt pro1}\={\tt pro1}\= {\tt pro1}\=\kill
  \> p ($\ols x$) \> \> $|$ \> (\ols t$_1$) \la (\ols x),\\[1mm]
  \> \> \> \>  Suc (\ols o$_{1,1}$) \la p$_{1,1}$(\ols i$_{1,1}$),\\
  \> \> \> \> $\vdots$\\
  \> \> \> \>  Suc(\ols o$_{1,k_1}$) \la p$_{1,k_1}$(\ols i$_{1,k_1}$),\\[1mm]
  \> \> \> \> let (\ols v$_{1,1}$) = q$_{1,1}$(\ols u$_{1,1}$),\\
  \> \> \> \> $\vdots$\\
  \> \> \> \> let (\ols v$_{1,l_1}$) = q$_{1,l_1}$(\ols u$_{1,l_1}$) \ 
  \ \ \ \= \\[1mm]
  \> \> \> \> \> =  \underline{Suc}(\ols s$_1$)\\[2mm]
  \> \> \> $|$ \> (\ols t$_2$) \la (\ols x),\\
  \> \> \> \> $\vdots$\\
  \> \> \> \> \> = \underline{Suc}(\ols s$_2$)\\[2mm]

  \> \> \> $\vdots$\\
  \> \> \> \underline{$|$ otherwise = Fail}
\end{tabbing}
} If one of the clauses of the above section is a unit clause, (i.e.~
$\tt p(\ol t, \ol s_j).$) or if its body contains no test predicates
then the corresponding line has the trivial guard \textsf{True}.  

Note that sequences may be empty, so if the predicate had no output
positions, then $\ols s$ would be the term $\sf ( \ )$.

\item If $\tt p$ is a \emph{non-test} predicate, then translation of
  the above section corresponds to the above script \emph{after
    removal} of the underlined parts; namely we have to eliminate from
  it the {\sf otherwise} statement and the \textsf{Suc}'s from the
  return values.  \HB
\end{itemize}
\end{definition}


Clearly, list constructions and built-in predicates need to be
handled separately, in particular, a test predicate of the form $\tt \olt t
== \olt s$ will be transformed to the term which returns either $\sf
Suc\ ()$ on success and $\sf Fail$ on failure.  We will abuse the
notation and also call this function $\sf ==$.



\begin{example} 
\label{exa:append}
Let us now consider the program {\tt append}. It is already plain, so,
assuming \texttt{append} to be non-test predicate, its translation is:

\begin{haskell} 
\> append (x1,x2)  
   \> $|$ \>  ([\ ], list) \la (x1, x2)\\
   \> \> \> \> = list\\[2mm]
 \> \> $|$ \> ((x:xs), list)  \la (x1, x2)\\
 \> \> $,$ \> let tail' = append (xs, list)\\
 \> \> \> \> = x:tail'
\end{haskell}
Notice  that nothing prohibits us from declaring \texttt{append}
as a \emph{test} predicate, if we do so we, the result of the
translation is:
\begin{haskell} 
\> append (x1,x2)  
   \> $|$ \>  ([\ ], list) \la (x1, x2)\\
 \> \> \> \> = Suc list\\[2mm]
 \> \> $|$ \> ((x:xs), list)  \la (x1, x2)\\
 \> \> $,$\> Suc tail' \la append (xs, list)\\
 \> \> \> \> = Suc (x:tail')\\[2mm]
 \> \> $|$ \> otherwise \ = Fail
\end{haskell}
In practice, the first program is more efficient than the second one
(though, this can differ per compiler), and it is more lazy than the
second one. This is further explained in the following \emph{aside}.
\II

\begin{remark} 
\label{rem:strictness}
The adopted partitioning has a natural influence on the strictness of
the resulting Haskell code. Consider the differences in the above
translations of \textsf{append}: if \textsf{append} is declared as
non-test then its translation will contain a let-expression in the guard \II

\textsf{let tail' = append (xs, list)\hfill(1)}
\II

\noindent
while if it is declared as test then, in its place, we will find the guard \II

\textsf{Suc tail' \la append (xs, list)\hfill(2)}
\II

\NI Now, while (1) is a let expression whose bound expression will
only be invoked if (and to the extent that) the value of
\textsf{tail'} is demanded, the second is a guard, which has to be
satisfied in order for the function it appears in to return a value.
Indeed, in (2) the term \textsf{append (xs,list)} will always be
reduced until it is completely computed, i.e.\ until it either reaches
either ``success'' or ``failure''. In this sense (2) is strict, while
(1) is lazy.

This behaviour is quite natural if one considers the following: since
test atoms might fail, we cannot trust their partial answers until we
have computed whether they'll succeed or not.  This implies that they
always have to be fully ``computed'', therefore forcing a strict
computation. On the other hand non-test predicates are guaranteed to
eventually succeed, so their computation might be stopped at the
moment that we have reached a partial result which is ``sufficient for
our purposes''. Therefore the non-test predicates naturally fit the
lazy model of computation.\HB
\end{remark}
\end{example}

\section{Logic Programs in a Lazy Functional Language}
We are at last in a position to demonstrate our thesis that the set of
logic programs considered as functional needs to be expanded.  As
issues, we consider logic variables, dynamic scheduling and
backtracking in turn.

The dynamics of some of the programs we are going to present in this
section is unavoidably rather complex, we apologize for the
inconvenience and ask the reader to resort to patience and
understanding.

\subsection{Logical Variables vs. Lazy Evaluation}
\label{sec:variables}

\emph{Logical variables} are one of the peculiarities of logic
programming. Most of the time, they are used in a standard way, that
is just as variables in an imperative language -- this is the case for
instance when the program is well-moded. Nevertheless there are many
important situations in which logical variables are exploited in all
their power. A typical such case is in the presence of
\emph{difference structures} such as \emph{difference lists}.

Here we show that even when used in a truly ``logical'' way, logical
variables are in many cases not an exclusive feature of logic
programs.

The following Polish Flag Problem example (incidentally, a simplified
version of Dijkstra's Dutch Flag Problem), reads as follows: given a
list of objects which are either red or white, rearrange it in such a
way that the red elements appear first and the white ones appear after
them. The following program is inspired by \cite[page 117]{O'K90}, we
have replaced ``$\setminus$'' by ``,'', thus splitting a position
filled in by a difference-list into two positions. Because of this
change in some relations, additional arguments are introduced.

\begin{program}
  \> polish(InList, RedWhites) \la \\
  \> \> distribute(InList, RedWhites, Whites, Whites,[ ]).\\[2mm]
  \> distribute([ ], Reds, Reds, Whites, Whites).\\
  \> distribute([X|Xs], [X|Reds0], Reds, Whites0, Whites) \la  red(X),\\
  \> \> distribute(Xs, Reds0, Reds, Whites0, Whites).\\
\> distribute([X|Xs], Reds0, Reds, [X|Whites0], Whites) \la  white(X),\\
\> \> distribute(Xs, Reds0, Reds, Whites0, Whites).\\[2mm]
\> mode polish(In, Out): non-test.\\
\> mode distribute(In, Out, In, Out, In): non-test.
\end{program}
Where we assume that predicates {\tt red} and {\tt white} are
appropriately defined elsewhere in the program and have mode
\texttt{red(In): test}. This program is plain and consistent, and by
translating it we obtain
\begin{haskell}
\> polish inlist 
 \> $|$ \> t \la inlist, \\
\>\>\> let (redwhites, whites) = distribute(t, whites, [\ ])\\
\>\>\>\> = redwhites\\[2mm]

\> distribute(is, rstail, wstail) 
\> $|$ \> ([\ ],rs,ws) \la (is,rstail,wstail)\\
\>\>\>  = (rs, ws)\\[2mm]
\>\> $|$ \>  (x:xs,rs,ws) \la (is,rstail,wstail)\\
\>\> , \> let (rshead, wshead) = distribute(xs, rs, ws)\\
\>\> , \> Suc () \la red x\\
\>\>\>\> =  (x:rshead, wshead)\\[2mm]
\>\> $|$ \>  (x:xs,rs,ws) \la (is,rstail,wstail)\\
\>\> , \> let (rshead, wshead) = distribute(xs, rs, ws)\\
\>\> , \> Suc () \la white x\\
\>\>\>\> =  (rshead, x:wshead)\\
\end{haskell}
Where \textsf{red} and \textsf{white} are again defined elsewhere in
the program.  This program runs perfectly well. Notice that the
definition of \textsf{polish} employs a \emph{circular data
  structure}: in fact the variable \textsf{whites} appears both on the
left hand side and on the right hand side of the expression $\sf
let \ (redwhites,whites) = distribute(t, whites, [\  ])$, in the
guard.

Circular data structures were first advocated by Bird \cite{Bird84} in
order to avoid multiple traversal of data structures, and since then
have become a standard tool of \emph{lazy} functional languages. It is
worth remarking that the above program (and most other programs
employing circular structures) would not function properly if we had
used a \emph{strict} functional language.

It is important to notice that the original logic program employs
logical variables in a highly non-trivial way. This is confirmed by
the fact that the program is not well-moded.

The fact that a program using difference-lists actually presents a
functional behaviour is not incidental. Consider an  atom
containing a difference-list $\tt \ldots p(t \backslash s)$ (for the
sake of simplicity, we assume that it does not have any other
argument), as above, we split this position in two, and obtain $\tt
\ldots p(t \backslash s)$. Now, the whole idea of having difference
lists, is that when a computation starting in (an instance of) $\tt
\ldots p(t \backslash s)$ will succeed, it will report a computed
answer substitution (c.a.s.) $\theta$ such that $\tt s\theta$ is
(``points to'') the tail of $\tt t\theta$. This implies that for all
$\sigma$ if $\tt s\theta\sigma$ is ground, then $\tt t\theta\sigma$ is
ground as well. Typically, after $\tt \ldots p(t \backslash s)$ has
succeeded with c.a.s.\ $\theta$, $\tt s\theta$ will eventually be
unified with a ground (classical) list (or with the head of another
difference-list, in which case the reasoning continues by considering
the tail of this second difference-structure). After this unification
has taken place, $\tt t\theta$ is going to be a classical list, which
can be employed as normal. In this sense we have that $\tt t$ depends
on $\tt s$, therefore $\tt t$ has to be considered output and $\tt s$
input, and the above atom should be translated into \textsf{let s =
  p(t)}; the only problem is that $\tt s$ is an input in disguise, in
the sense that when $\tt p(t,s)$ is called, $\tt s$ is typically not
(yet) ground (in other words, it is not yet known). However, this is
hardly a problem when we consider \emph{lazy} functional languages (at
the same time, it is the reason why programs with difference lists
cannot be easily translated into a strict functional language).

One could argue that the above reasoning could be completely reversed,
starting by saying that ``$\ldots$ for all $\sigma$ if $\tt
t\theta\sigma$ is ground, then $\tt s\theta\sigma$ is ground as well,
$\ldots$, thus if $\tt t\theta$ is unified to a ground list, then $\tt
s\theta$ will become a ground list, and this shows that $\tt s$
depends on $\tt t$, and that therefore the above atom should be
translated into \textsf{s = p(t)}'' (which would fail to function).
This is in principle true (dependencies in LP are always
bidirectional), however, this property is never used, indeed it cannot
be used in practice for the following simple reason: after succeeding
with c.a.s.  $\theta$, we typically have that $\tt t\theta \equiv
[a_1,a_2,\ldots,a_k | X]$ and that $\tt s\theta \equiv X$. Now, while
it is always possible to unify $\tt s\theta$ with any ground list $\tt
l$, trying to do this with $\tt t\theta$ will almost certainly lead to
failure (unless $\tt [a_1,a_2,\ldots,a_k]$ is a prefix of $\tt l$).
Therefore difference-lists are virtually always employed in a
\emph{directional} fashion.

Another example of a program using logical variables which can be
safely translated into Haskell is given in the following section. Of
course, one can find an example of a program using difference-lists
which would not work in Haskell; actually, counterexamples are
extremely easy to contrive: variables in LP are always adirectional,
and if one fully exploits this will always obtain a program which has
no functioning functional counterpart. We don't want to deny this, on the
contrary: here we are interested in how programs are \emph{usually}
used, and in pointing out that some standard methodologies which are
normally considered as applicable only to LP, are actually not
so.

Of course not all programs using logical variables are translated
correctly: typical such examples are the programs which incrementally
fill in a data structures such as in the \emph{eight queen} example
and in the \texttt{SEQUENCE} example in Appendix
\ref{sec:sequence} (these programs use unification in a crucial way,
and this is confirmed by the fact that they are not
\emph{consistent}). Other examples are circular programs such as the
following one
\begin{program}
  \> p(X) \la eq(X, X).\\
  \> eq(X, X).
\end{program}
moded as follows: \texttt{p(In:Ground):non-test} and
\texttt{eq(Out:Ground, Out:Ground):non-test}. This program is circular
in a non-well-founded way, and this, when translated, yields a program
which is not \emph{productive}.

We can safely conclude that difference lists have a natural
counterpart in the circular structures of lazy functional programming.


\subsection{Dynamic scheduling vs.\ Lazy Evaluation}
\label{sec:scheduling}
Another prominent property of logic programming is the possibility of
having a dynamic selection rule, possibly guided by appropriate
\emph{delay declarations}. Let us consider the following example,
which, given the list {\tt Xs} of integer values, {\tt
  del\_max(Xs,Zs)} produces the list {\tt Zs} by deleting all the
occurrences of its maximum element.

\begin{program}
\> del\_max(Xs, Zs) \la find\_max\_and\_del(Xs,  Max, Zs, Max).\\[2mm]
\%\> find\_max\_and\_del(InList,El,OutList,Max) \\
\%\> \> Max {\rm is the maximum element of the list }InList, {\rm and}\\
\%\> \> OutList {\rm is obtained from }Inlist {\rm by deleting all the
  occurrences of } El {\rm from it}\\[1mm]
\> find\_max\_and\_del([ ],  \_, [ ], 0).\\
\> find\_max\_and\_del([X | Xs],  El, Ys, Max) \la \\
\>   \>  find\_max\_and\_del(Xs, El, Zs, Max'),\\
\>   \>  sup(X, Max', Max), \\
\>   \>  del\_if\_first([X | Zs], El, Ys).\\[2mm]
%
\> del\_if\_first([EL | Zs], El, Zs).\\
\> del\_if\_first([X | Zs], El, [X|Zs]) \la X $\neq$ El.\\[2mm]
\> mode del\_max(In:List[Int], Out:List[Int]): non-test.\\
\> mode find\_max\_and\_del(In:List[Int],In:Int,Out:List[Int],Out:Int):non-test.\\
\> mode sup(In:Int, In:Int, Out:Int): non-test. \% {\rm defined in the
  obvious way}\\
\> mode del\_if\_first(In:List[Int], In:Int, Out:List[Int]): non-test.
\end{program}

\NI It is worth noticing that the program uses logical variables in a
nontrivial way. This is confirmed by the fact that it is not
well-moded.  Specifically, the variable \texttt{Max} in the first
clause is used as an asynchronous communication channel between
processes, as the atom {\tt find\_max\_and\_del(Xs, Max, Max, Zs)}
uses \texttt{Max} as input value that it has to produce itself. 

Furthermore, the program requires an appropriate \emph{dynamic
  scheduling}.  In fact, when run with a standard left-to-right
selection rule, the query {\tt del\_max(ts, Zs)} ($\tt ts$ being a
list of natural numbers) leads to a run-time error (or to an incorrect
answer), and, provided that we fix this problem, to a very inefficient
computation.

The first problem (concerning the runtime error) is due to the fact
that the computation will soon a goal of the form
\texttt{del\_if\_first(ts, El, Zs)}, where \texttt{ns} $\tt (=
[n|ns'])$ is a non-empty list of integers, and \texttt{El} and
\texttt{Zs} are distinct variables. At that point the interpreter will
proceed and might reach the call $\tt n \neq El$, which -- being
\texttt{El} a variable -- will flounder\footnote{In practice, the
  behaviour of the interpreters in these situations are different,
  this depends on the non-logical behaviour of the built-ins of PROLOG
  and on whether we employ \texttt{==} in order to make the program
  plain and $\backslash$\texttt{==} in order to implement $\neq$. For
  instance in Eclipse and SWI-Prolog the call \texttt{n == El} fails,
  and a subsequent call \texttt{n $\backslash$== El} succeeds (!) in
  Eclipse by returning the empty c.a.s.\ and in SWI-prolog by
  instantiating \texttt{El} to an apparently random numeric value. Of
  course, this behaviours lead to solutions which are almost always
  incorrect.}. In other words, this program cannot be run with the
normal leftmost selection rule.

The second problem (concerning program's inefficiency) is due to the
fact that the query {\tt del\_max(Xs, Zs)} could return the list $\tt
Zs$ in linear time (scanning $\tt Xs$ only once), however, it is easy
to see that if we employ any fixed selection rule, the program has to
go through a remarkable amount of backtracking, which makes it run in
quadratic time on the length of the input list\footnote{Informally,
  the reason is that the interpreter will run into some calls of the
  form $\tt del\_if\_first( \ldots , El, Zs)$, where $\tt El$ is still
  a variable. For this reason $\tt del\_if\_first( \ldots , El, Zs)$
  will take a guess and delete an element which will usually turn out
  not to be the correct one, and this will eventually produce
  backtracking.  Since the number of needed backtracking steps is
  linear in the size of the input list, this decreases the performance
  of the program from linear to quadratic time. In order to avoid this
  problem, we have to make sure that an atom of the form $\tt
  del\_if\_first(\_, El, \_)$ will be selected in the derivation {\em
    only when\/} $\tt El$ will be instantiated to a ground term. We
  can do this by employing the above delay declarations.}.

Both problems can be solved by employing a \emph{dynamic selection
  rule} and by prohibiting the selection of certain atoms until their
arguments are sufficiently instantiated using for instance the
following \emph{delay declarations} \cite{Nai82}:
\begin{program}
\> delay sup(X, Y, \_) until ground(X) $\tt \A$   ground(Y).\\
\> delay $\neq$(X,Y) until ground(X) $\tt \A$   ground(Y).\\
\>  delay del\_if\_first([X|Xs], El, \_) until ground(X) $\tt \A$
  ground(El)
\end{program}
For instance, the first declaration will suspend any call to
\texttt{sup(t, s, v).} until \texttt{t} and \texttt{s} are ground
terms.  Delay declarations have become an important standard control
tool and are implemented in various versions of Prolog (for instance
in Sixtus Prolog and in Eclipse \cite{WV93}) and in the language
G\"odel \cite{HL94}.

Now, let us for a moment not bother about the delay declarations and
translate this program into Haskell. We obtain the following script.


\begin{haskell}
\> del\_max as \> $|$ \> xs \la as\\
\> \>              ,   \> let (zs, maxel) = find\_max\_and\_del (xs,maxel)\\
\>\>\>\>  = zs\\[2mm]

\> find\_max\_and\_del (x1,x2)  
        \> $|$  \> ([ ], el) \la (x1, x2)\\
\>\>\>\>        =  ([ ], 0)\\[2mm] 
\>      \> $|$  \> (x:xs, el) \la (x1,x2)\\
\>              \> , \> let       (zs, maxel') = find\_max\_and\_del (xs,el)\\
\>              \> , \> let       maxel = max x maxel'\\
\>              \> , \> let       ys = del\_if\_first (x:zs, el) \\
\>\>\>\>                = (ys, maxel) \\[2mm]

\> del\_if\_first (x1, x2) 
        \> $|$  \> ([ ], el) \la (x1, x2)\\
\> \>\>\>                               = [ ] \\[2mm]
\>      \> $|$  \> (x:xs, el) \la (x1,x2)\\
\>\> , \>          x == el\\
\> \>\>\>                               = zs\\[2mm]
\>      \> $|$  \> (x:xs, el) \la (x1,x2)\\
\>\> , \>          x /= el\\
\> \>\>\>                               = x:zs
\end{haskell}
This program works fine, and his runtime complexity is linear in the
size of the input. We can therefore state that the lazy computational
mechanism compensates for the lack of control over dynamic scheduling,
without which the above logic program could not be run or would have a
quadratic complexity.

Thus although the mechanism of lazy evaluation and delay declarations
are quite different (actually, they are the opposite: the call-by-need
mechanism determines which term \emph{has} to be reduced, while delay
declarations determine which atoms \emph{should not} be resolved),
they often accomplish the same thing.


The fact that lazy evaluation here plays a crucial role is confirmed
by the fact that, if we had declared all predicates to be test
predicates (thus forcing strictness, as explained in the 
Remark \ref{rem:strictness})
the translated program would not function properly.  


Thus again we are in presence of a program exploiting logical
variables in a complex way which nevertheless has a natural
translation into Haskell.

\subsection{Backtracking and Nondeterminism}

Another outstanding feature of logic programs is their backtracking
mechanism, which virtually implements a \emph{don't know}
nondeterministic system.

In the light of the above examples, we believe that nondeterminism is
by far the most important and the mostly used peculiar feature of the
logic programming paradigm. We don't want to challenge this, on the
contrary. At the same time, it is important for us to show to which
extent  a (lazy) functional program can mimic a logic program which
uses backtracking.


Consider the following program.

\begin{program}
\> backtracker(X) :- producer\_a(Y), picky\_modifier(Y,X).\\
\> backtracker(X) :- producer\_b(Y), picky\_modifier(Y,X).\\
\> producer\_a("a").\\
\> producer\_b("b").\\
\> picky\_modifier("b", "c").
\end{program}
The adopted mode and partitioning is 
\begin{program}
\> backtracker(Out) : non-test \\ 
\> producer\_a(Out) : non-test {\rm ( and the same for }
producer\_b \textrm{)}\\
\> picky\_modifier(In, Out) : test
\end{program}
Its translation is the following:
\begin{haskell}
\> backtracker 
\>      $|$     \> let y = producer\_a\\
\> \> , \>                Suc x \la picky\_modifier y\\
\> \> \> \>             = x\\[2mm]

\> \>   $|$     \> let y = producer\_b\\
\> \> , \>                Suc x \la picky\_modifier y\\
\> \> \> \>             = x\\[2mm]

\> producer\_a = "a"\\
\> producer\_b = "b"\\[2mm]

\> picky\_modifier     x        
\>      $|$ \> x == "b" \\
\> \> \> \> = Suc  "c"\\[2mm]
\> \>   $|$ \> otherwise \ = Fail\\
\end{haskell}


The Haskell translation is able to report all the correct answers,
even though in LP for the query $\tt \la backtracker(X)$ in order to
return the answer {\tt X = "a"}, the interpreter has to go through
some backtracking. Notice in fact that the above logic program is not
deterministic.

Consider now the following program scheme:

\begin{program}
  \> p(X) \la generate(X), test(X).
\end{program}
It is immediate to translate it and to check that if \texttt{generate}
has more than one solution then the translation does not behave as the
logic program does: while the query {\tt :- p(X)} succeeds provided
that one of the solutions of \texttt{generate(X)} satisfies
\texttt{test(X)}, the Haskell translation manages to report one answer
only so in the unlikely case that the first solution founded by
\texttt{generate(X)} satisfies \texttt{test(X)}; in all other cases
$\sf p$ reduces to $\sf Fail$.

The key factor for the translation to work correctly we need to avoid
logic programs in which consistent queries might originate SLD trees
with more than one successful (sub-) branch.  There exists techniques
based on list-comprehension in order to translate logic programs into
functional programs in such a way that the resulting program will
(eventually, lazily) report the list of all the answers that the
initial logic program would. In those cases, however, one can clearly
not talk of a \emph{literal} translation, which is the starting point
of our research (programs able to return more than one answer are in
our opinion intrinsically logic programs, and therefore do not belong
to our target).

To be precise, a non-deterministic logic program can be safely
translated onto Haskell provided it is \emph{input discriminative}, as
defined as follows:

\begin{definition}[Input Discriminative]
\label{def:id2}
  Let $\PP$ be a program, $M_{\PP}$ be its least Herbrand model,
  and
\begin{program}
\> p$_1$(\olt i$_1$, \olt o$_1$) \= \la \olt{test}$_1$, \olt{rest}$_1$.\\
\> \> \ $\vdots$\\
\> p$_n$(\olt i$_n$, \olt o$_n$) \la \olt{test}$_n$, \olt{rest}$_n$.
\end{program}
Be the complete set of the rules of $\PP$ (where the conjunction
\olt{test}$_i$ contains only test predicates and \olt{rest}$_i$
contains only non-test predicates). We say that $\PP$ is \emph{input
  discriminative} if for each $j \neq k \in [1,n]$, such that
$\mathtt{p}_j = \mathtt{p}_k$ we have that
\begin{itemize}
\item
for each ground $\theta$ such that $\olt i_j\theta = \olt i_k\theta$
we have that $M_{\PP} \models \neg \ (\olt{test}_j\theta \A
\olt{test}_k\theta)$.
\HB
\end{itemize}
\end{definition}

It is worth noticing that this concept of input discriminative program
is rather less restrictive than the concept of \emph{deterministic}
program, and that input-discrimitative programs might still require
non-trivial (non-shallow, see \cite[Ch.~6]{SS86}) backtracking. We
could say that these programs admit some \emph{shallow
  nondeterminism}.

Summarizing, there is a point to be remarked, that -- strictly
speaking -- the Haskell translation of a program can \emph{always}
mimic the backtracking taking place in the original logic program.
What the Haskell translation can\emph{not} do is report multiple
answers.

\subsubsection*{Failure, Nondeterminism and Related Work}
\label{sec:multiple}
The feature of logic programs of being able of reporting more than
answer, and how this is handled in the different translation systems
is a topic which deserves a separate discussion.

Regarding this issue, the literature on papers presenting a
translation from logic to functional programs can be divided in two
main groups.

On one side we find papers which are not concerned with the
nondeterminism (or the backtracking) mechanism of logic languages
\cite{Mar94-trs,GW92,kRKS98,vRaa97}, these papers are usually mainly
concerned in providing a transformation system which allows one to
prove program properties such as termination of the original logic
program. For this reason they focus on obtaining a translation which
maps only the non-failing computations correctly. In these papers the
failure and backtracking mechanism are disregarded during the
translation.

On the other side, we find \cite{Mar95,Red84}, in which the authors
propose a translation in which the full (PROLOG-like) computational
mechanism is preserved, including the possibility of having multiple
answer for the same query and the possibility of failure. This is
achieved by letting a query return the \emph{list} of computed
answer substitutions, where the empty list corresponds to the failing
case, in the same way advocated by Wadler \cite{Wad85}.  The lazy
computational mechanism then takes care of computing only those
answers which are necessary, and backtracking is faithfully rendered
by a standard list-comprehension schema.

The translation system we have employed lies somewhere in the middle
between those two methods.  Our goal was to take also failure into
account, yet retaining a literal translation system, in which the
computational mechanism of the resulting functional program is as
similar as possible to the one of the original logic program.

Of course we can only correctly translate programs which do not return
more than one answer for the same query (at the same time, it is
important to notice that these programs don't have to be
deterministic; for instance \texttt{member} is nondeterministic).

In our opinion, the possibility of returning more than one answer is
to be considered a \emph{peculiar} one of the LP paradigm, and the
fact that it can be emulated by functional programs does not
obliterate our position.

\section{Conclusions}

The goal of our research was to investigate to which extent some
features considered peculiar of the logic programming paradigm are
really so.  For this purpose we have devised a simple -- literal --
system which enabled us to translate logic programs into the lazy
functional language Haskell.

It is known (see also \cite{Mar94-trs,GW92,kRKS98,vRaa97}) that
\emph{if we restrict our attention to non-failing, non-backtracking
  computation} then well-moded simply moded programs have a natural
counterpart in a functional language.  The properties of being
well-moded and simply moded indicates a manner of use of variables in
logic programming which is undoubtedly ``functional''.  To this
statement we want to add that well- and simply moded programs can be
considered as \emph{strictly} functional, as they can be safely
translated into a \emph{strict} functional language.

In this paper we have shown that in a \emph{lazy} functional language,
this picture broadens significantly, and some of the features that
were -- in the light of the results above -- commonly considered as
exclusive of the logic programming paradigm, can be naturally found in
a lazy functional language such as Haskell. 

In particular, we have shown that the use of complex logical variables
in data structures such as difference lists (or such as in 
program in \texttt{del\_max})
find a natural counterpart in the circular structures \cite{Bird84} of
lazy functional programs. These structures that were commonly
considered as ``structurally logical'' are thus not so.  We can then
attempt a rough classification of logic programs according to the
level of complexity at which they employs their variables
(backtracking and nondeterminism is not considered here). We then have
the following division.
\begin{enumerate}\smallromani

\item \emph{Strictly Functional} programs which use variables in a standard
  (imperative-like) way. These are characterized by being
  \emph{well-moded} (or by being so after permutation of the clause's
  body atoms). 

\item \emph{Lazy Functional} programs which admit a safe translation
  into Haskell: i.e.\ programs which can be translated into Haskell
  (via the syntactic translation) and whose operational behaviour is
  isomorphic to the one of their functional counterpart. 

\item \emph{Intrinsically Logical} programs which do not admit a safe
  translation into Haskell with our translation scheme.
\end{enumerate}
This raises the interesting question of how large is the class of
intrinsically logical programs. Without pretending to be able to
characterize extensively this limit, it is interesting to notice that
programs which are plain and consistent and which either admit a
Layered Mode \cite{EG96-workshop}, or are S-well-typed programs
\cite{BM97} are safely translatable into Haskell (modulo the
possibility of backtracking, which is discussed in the sequel). As
argued in \cite{EG96-workshop}, we believe that these programs
actually encompass the majority of actual programs which use logical
variables in a non-elementary way.  We think that a classification and
understanding of these levels might be useful both to enhance the
performance of logic languages (as already done to some extent in the
language Mercury) and to prove more precise program
properties.  \II

Furthermore, we have also addressed another logical feature: the
possibility of dynamic scheduling. In theory in LP \emph{any} atom is
selectable as all selection rules yield the same successful
derivations. In practice this does not work, and adopting a random
selection rule would in the best case yield to an explosion of the
search space; for this reason PROLOG uses a fixed left-to-right
selection rule, a feature which is either explicitly or implicitly
always exploited by the programmers. However, some programs (like
\texttt{delmax} above, or concurrent-like programs) are not correct
under a fixed search rule. In these cases the ``right'' selection
strategy is enforced by the use of appropriate delay declarations
(d.d.), which serve to indicate which atoms in a query should not be
resolved.  The implementation of d.d.~is rather costly, as atoms are
continuously being suspended and forced.  Here we have seen one
example in which the lazy evaluation mechanism of Haskell achieves the
same effect of the use of d.d.. As we have pointed out, call-by-need
can be regarded as a dynamic selection strategy, which is however
based on an principle opposite to the one of d.d.~in the sense that
call-by need determines which term has to be reduced, while delay
declarations determine which atoms should \emph{not} be resolved. A
naturally arising question here is whether it is possible to implement
in logic programming languages a selection rule which is ``driven'' by
a call-by-need mechanism, instead of ``restricted'' by the use of
delay declarations. This could possibly lead to reduction of the
suspension overhead and thus to performance improvements. The
difficulty in implementing such a search rule lies in the fact that in
LP it is not clear which output values depend on which input values
(actually, it is not clear what is input and what it is output to start
with), so in order to implement such an intelligent selection strategy,
one would need some sophisticated analysis tools which might either be
based on abstract interpretation (with tools similar to the ones of
\cite{CDG93}), or on refined versions of modes such as the ones
described in \cite{BM97,EG96-workshop}. Other works related to this
subjects are \cite{Lut92,EvR98}.  \II

We have also discussed the fact that logic programs allow
backtracking. We have seen that -- strictly speaking -- backtracking
computations can be easily mimicked by the functional language by an
appropriate use of the guards; what cannot be (easily) mimicked in
Haskell is the possibility of returning multiple answers, at least not
unless one uses additional constructs such as the list-of-successes
method \cite{Wad85}.  An interesting research direction might be to
define appropriate monadic structures (such as those in \cite{Wad92})
to capture the failure or success and returning of multiple arguments.
This would broaden the set of logical programs which we can capture
with our simple translation scheme, without adding signifant
complexity to it.
\II

In conclusion, we have demonstrated with a simple, literal translation
scheme that several features considered as belonging specifically to
logic programming are found naturally in lazy functional programming,
dismissing the folklore that the functional core of logic programming
is contained in the set of well- and simple moded programs.

\appendix

\section{Well-Moded Programs}

The following concept is essentially due to Dembinski and Maluszynski
\cite{DM85}; we use here an elegant formulation due to Rosenblueth
\cite{Ros91}.  

\begin{definition} \label{def:wm} 
  A clause $\tt p_0(\ol t_0,\ol s_{n+1}) \la p_1(\ol s_1,\ol t_1), \LL,
  p_n(\ol s_n,\ol t_n)$ is called \emph{well-moded} if for $i
  \in [1,n+1]$
$$\tt \var(\ol s_i) \subseteq  \bigcup_{j=0}^{i-1} \var(\ol t_j).$$

\NI A query $\tt \ol A$ is called \emph{well-moded} iff the clause
$\tt q \la \ol A$ is, where $\tt q$ is any (dummy) atom of zero arity.
\II
 
\NI A program is called \emph{well-moded} if every clause of it is.
\HB
\end{definition}
 
It is important to notice that the first atom of a well-moded goal is
ground in its input positions and a variant of a well-moded clause is
well-moded.  Furthermore, the notion of of well-modedness, is
``persistent'', as shown by the following Lemma. Recall that a
LD-resolvent is a resolvent in which the leftmost atom per the query
is the selected one, and that an LD-derivation is a derivation
obtained employing the leftmost selection rule, 
 
\begin{lemma} \label{lem:wm-persist}
  An LD-resolvent of a well-moded goal and a well-moded clause that is
  variable-disjoint with it, is well-moded.  \HB
\end{lemma}

The next result is originally due to Dembinski and Maluszynski and
follows directly from the definition of well-moded program.  
 
\begin{corollary}\label{cor:wm}
  Let $\tt P$ and $\tt \ol A$ be well-moded, and let $\xi$ be an LD-derivation
  of $\tt \ol A$ in $\tt P$.  All atoms selected in $\xi$ contain ground terms
  in their input positions.  \HB
\end{corollary}

That is, in presence of well-moded programs and queries, if we use a
left-to-right computation schema we are sure that every time that we
select an atom, the ``value'' of his input arguments has already been fully
computed. This shows that well-moded programs have a straightforward
left-to-right data-flow.

Under certain conditions well-moded programs are also
\emph{unification-free}. To show this, we need a definition first. The
following notion was first defined in \cite{AE93}.
 
\begin{definition} \label{def:sm} 
  A clause $\tt p_0(\ol s_0,\ol t_{n+1}) \la p_1(\ol s_1,\ol t_1), \LL,
  p_n(\ol s_n,\ol t_n)$ is called \emph{simply moded} if if $\tt \ol t_1,
  \dots , \ol t_n$ is a linear family of variables and for $i \in
  [1,n]$
\[
\var(\tt \ol t_i)\cap (\bigcup_{j=0}^{i} \var(\tt \ol s_j))=\ES.
\]
 
\NI A query $\tt \ol A$ is called \emph{simply moded} iff the clause $\tt q
\la \ol A$ is, where $\tt q$ is any (dummy) atom of zero arity.  \II
 
\NI
A program is called {\em simply moded\/} if every clause of it
  is.  \HB
\end{definition}
 
Thus, assuming that in every atom the input positions occur first, a
clause is simply moded if all output positions of every body atom are
filled in by distinct variables, which do not occur earlier in the body nor in
an input position of the head.
 
It is worth noticing that -- as shown by the little survey in
\cite{AE93} -- most programs are already simply-moded and that often
non simply-moded programs can naturally be transformed into
simply-moded ones, for instance the non-simply-moded clause $\tt
last(List,El) :- reverse(List,[El|\_]).$ can be transformed into $\tt
last(List,El) :- reverse(List,List'), [El|\_] = List'$.

The property of being simply moded is also ``persistent'' in the sense
that the resolvent of a simply moded query with a simply moded clause
is simply moded.

In \cite{AE93} it is proven that if the program and the query are
simply moded, then they generate an LD-derivation which is
\emph{unification-free}, i.e.\ that each time an atom $\tt A$ is
selected and resolved in it via a clause $\tt H \la \ol B$, then the
unification of $\tt A$ and $\tt H$ does not really require a full
unification algorithm, but can always be reduced to a \emph{double
  matching:} one (``from'' $\tt A$ ``to'' $\tt H$) for the input
positions and a second one (``from'' $\tt H$ ``to'' $\tt A$) for the
output ones. This result clearly shows that simply
and well-moded logic programs are functional in nature (besides for
the possibility of reporting multiple answers, of course).

\section{More on Haskell}
\label{app:more}

The following (nonsense) program embodies most of the concepts we use:

\begin{haskell} 
\> append ([\ ],x) \> = Suc x\\
\> append (x1,x2)  
  \> $|$ \> (x:xs) \la x1\\
 \> \> $,$\> Suc tail \la append (xs, x2)\\
 \> \> $,$\> let newtail = (x:tail)\\
 \> \> \> \> = Suc newtail\\
 \> \> $|$ \> otherwise \ = Fail
\end{haskell}

In a call of \textsf{append (a,b)}, the first equation will be tried
first.  Here, \textsf{a} will be pattern matched to the empty list.
If this succeeds, then \textsf{x} is matched to \textsf{b} (since both
are variables, this will always succeed), and finally \textsf{Suc x}
is returned.  If the pattern match above failed, then the first guard
will be tried, which tries to pattern match the first element of the
tuple to a list with at least one element.  Should this succeed, the
result of a recursive call to \textsf{append} is matched against
\textsf{Suc tail}, and if successful \textsf{(x:tail)} is bound to the
variable $\sf newtail$, followed by the returning of $\sf Suc\ 
newtail$.  If either of the pattern matches failed, then the second
guard will be tried.

\section{program \texttt{SEQUENCE}}
\label{sec:sequence}

This example is provided by the Prolog formalization of a problem from
Coelho and Cotta \cite[pag. 193]{CC88}: arrange three 1's, three 2's,
..., three 9's in sequence so that for all $i \in [1,9]$ there are
exactly $i$ numbers between successive occurrences of $i$.
\begin{program}
\>  sublist(Xs, Ys)  \la \ {\tt Xs} {\rm is a sublist of the list} {\tt
Ys}. \\
\> sublist(Xs, Ys) \la \ app(\_, Zs, Ys), app(Xs, \_, Zs).  \\[2mm]
\> sequence(Xs) \la \  {\rm {\tt Xs} is a list of 27 elements.} \\
\> sequence([\_,\_,\_,\_,\_,\_,\_,\_,\_,\_,\_,\_,\_,\_,\_,\_,\_,\_,\_,\_,\_,\_,\_,\_,\_,\_,\_])
. \\[2mm]
\> question(Ss) \la \  {\rm {\tt Ss} is a list of 27 elements forming the desired sequence.} \\
\> question(Ss) \la \\
\>   \> sequence(Ss), \\ 
\>   \> sublist([1,\_,1,\_,1], Ss), \\ 
\>   \> sublist([2,\_,\_,2,\_,\_,2], Ss), \\ 
\>   \> sublist([3,\_,\_,\_,3,\_,\_,\_,3], Ss), \\ 
\>   \> sublist([4,\_,\_,\_,\_,4,\_,\_,\_,\_,4], Ss), \\ 
\>   \> sublist([5,\_,\_,\_,\_,\_,5,\_,\_,\_,\_,\_,5], Ss), \\ 
\>   \> sublist([6,\_,\_,\_,\_,\_,\_,6,\_,\_,\_,\_,\_,\_,6], Ss), \\ 
\>   \> sublist([7,\_,\_,\_,\_,\_,\_,\_,7,\_,\_,\_,\_,\_,\_,\_,7], Ss), \\ 
\>   \> sublist([8,\_,\_,\_,\_,\_,\_,\_,\_,8,\_,\_,\_,\_,\_,\_,\_,\_,8], Ss), \\ 
\>   \> sublist([9,\_,\_,\_,\_,\_,\_,\_,\_,\_,9,\_,\_,\_,\_,\_,\_,\_,\_,\_,9], Ss). \\[2mm]
\>   {\rm augmented by the {\tt append} program}.
\end{program}

\end{document}